\documentclass[prd,aps,floatfix,nofootinbib,preprint,tightenlines ]{revtex4}
\usepackage{dcolumn}% Align table columns on decimal point
\usepackage{amsmath}
\usepackage{latexsym}
\usepackage{graphicx}
\usepackage{bm}
\newcommand{\bee}{\begin{equation}}
\newcommand{\ee}{\end{equation}}
\newcommand{\beea}{\begin{eqnarray}}
\newcommand{\eea}{\end{eqnarray}}

\begin{document}
%%%%%%%%%%%%%%%%%%%%%%%%%%%%%%%%%%%%%%%%%%%%%%%%%%%%%%%%%%%%%%%%%%%%%%
\title{
Finite-size scaling tests for spectra in SU(3) lattice gauge theory coupled to 12 fundamental flavor fermions}
\author{Thomas DeGrand}
\email{thomas.degrand@colorado.edu}
\affiliation{Department of Physics,
University of Colorado, Boulder, CO 80309, USA }

\begin{abstract}
I carry out a finite-size scaling study of the correlation length in SU(3) lattice gauge theory
coupled to 12 fundamental flavor fermions, using recent data published by
Fodor, Holland, Kuti, N\'ogradi and Schroeder \cite{Fodor:2011tu}. I make the  assumption that the
system is conformal in the zero-mass, infinite volume limit,
that scaling is violated by both nonzero fermion mass and by finite volume, and that the
 scaling function in each channel is determined self-consistently by the data.
From several different observables I
extract a common exponent for the scaling of the correlation length $\xi$ with the fermion mass
$m_q$, $\xi \sim m_q^{-{1/y_m}}$ with
$y_m \sim 1.35$. Shortcomings of the analysis are discussed.
\end{abstract}

%\pacs{11.15.Ha,  12.60.Nz}
\maketitle

A recent paper by Fodor, et al  \cite{Fodor:2011tu} presents an analysis of
spectroscopy for $SU(3)$ gauge theory coupled to $N_f=12$ flavors of
fermions. This theory is a potential candidate for
 beyond - Standard Model physics. However, it is a subject of
some recent controversy. Appelquist et al \cite{Appelquist:2007hu,Appelquist:2009ty},
 performing a calculation of a running coupling constant, concluded that
in the limit of vanishing fermion mass,
it had an infrared-attractive
fixed point (IRFP). In that limit it exhibits conformal behavior at long distances.
The authors of Ref.~\cite{Fodor:2011tu} collected spectroscopic
data at one value of the gauge coupling,
four simulation volumes, and eight fermion masses, for a total of twelve
volume - mass combinations. They analyzed their data under the competing assumptions that the
system was confining and chirally broken, or conformal,
 and concluded that their data favored the confining and chirally broken scenario.
Other references relevant to this controversy include
\cite{Deuzeman:2008sc,Fodor:2009wk,Deuzeman:2009mh,Hasenfratz:2009ea,Jin:2009mc,Hasenfratz:2009kz,Hasenfratz:2010fi,Deuzeman:2010fn,Deuzeman:2010gb,Hasenfratz:2011xn}.

The ``conformal scenario'' for a system like the one under discussion
assumes that the long distance behavior of the theory is described by
one relevant coupling, the fermion mass $m_q$. In infinite volume, tuning
 the mass to zero causes the correlation length to diverge algebraically,
\bee
\xi \sim m_q^{-\frac{1}{y_m}}.
\label{eq:corrlen}
\ee
The quantity $y_m$ is the leading relevant exponent for the system, in statistical physics language.
This exponent is related to the anomalous dimension $\gamma_m$ of the mass operator $\bar\psi\psi$,
and determines the running of the mass parameter according to
\bee
\mu \frac{dm(\mu)}{d\mu} = -\gamma_m(g^2) m(\mu),
\ee
$y_m = 1+\gamma_m(g_*)$.
All other couplings, including the gauge coupling (more properly, the distance of the gauge coupling
from its fixed point value) are irrelevant.

However, no simulation is ever done in infinite volume. The system size $L$ is also a relevant parameter since
the correlation length only diverges in the $1/L\rightarrow 0$ limit.
When the correlation length measured in a system of size $L$
(call it $\xi_L$) becomes comparable to $L$, $\xi_L$ saturates at $L$ even as $m_q$
vanishes. 
However, if the only large length scales in the problem are $\xi$ and $L$,
then overall factors of length can only involve $\xi$ and $L$. For the correlation
 length itself,
this argument says that
\bee
\xi_L = L F(\xi/L)
\label{eq:fss1}
\ee
where $F(x)$ is some unknown function of $\xi/L$. A somewhat more useful version of this
relation invokes Eq.~\ref{eq:corrlen}, to say
\bee
\xi_L = L f(L^{y_m} m_q)   .
\label{eq:fss2}
\ee
Then one can plot $\xi_L/L$ vs $L^{y_m} m_q$ for many $L$'s, and vary $y_m$. Under this variation,
 data from different $L$'s will march across the x axis
at different rates. The exponent can be determined by tuning $y_m$ to collapse
the data onto a single curve.

The authors of Ref.~\cite{Fodor:2011tu} confronted a subset of their data with Eq.~\ref{eq:corrlen}
and all their data with Eq.~\ref{eq:fss1}. The reason for this note and my analysis is that they
chose
a particular functional form for $F(x)$ in their fit. However, in general,
 the actual form of the scaling function is unknown, Limiting behavior is known:
for example, at large $x$, $F(x)=x$ and at small $x$, $F(x)$ goes to a constant.
This means that finding $F(x)$ or $f(x)$ is itself part of the fit.
In addition, there is no reason for different observables to have the same $F(x)$. The finite
box size can -- and the data show that it does -- affect them differently.

Not having a-priori knowledge of the scaling function means that it is
difficult to assign a goodness-of-fit parameter, like a chi-squared, to a determination
of $y_m$.
 All one can do is to compare the $y_m$'s from different data sets, and ask if they are consistent.
It also means that the resulting value of $y_m$ will have a large uncertainty.
However, I think it is still a potentially informative task, to ask, whether the 
data of Ref.~\cite{Fodor:2011tu}
is consistent with the finite size scaling hypothesis, while letting the data itself
determine the scaling function. That is the subject of this note.

This methodology was used in Ref.~\cite{DeGrand:2009hu} to measure $y_m(g^2)$ in the
$SU(3)$ - $N_f=2$ sextet system. It produced relatively noisy exponents.
The technique of using correlation functions in the Schr\"odinger functional
is much more accurate, but the finite-size scaling exponents agreed reasonably well with
these better measurements \cite{DeGrand:2010na}.

The technique has already been described in  Ref.~\cite{DeGrand:2009hu}, so we will just
proceed to results. I have analyzed the mass spectra of the
following states from the data sets of
Ref.~\cite{Fodor:2011tu}: the pseudoscalar (would-be Goldstone boson in a chirally broken theory),
vector and axial vector mesons, baryon, and pseudoscalar decay constant, $f_\pi$.
I will define the correlation length $\xi_L$ to be just the inverse mass, (or $1/f_\pi$) in 
a lattice whose spatial length is $L$.

\begin{figure}
\begin{center}
\includegraphics[width=0.6\textwidth,clip]{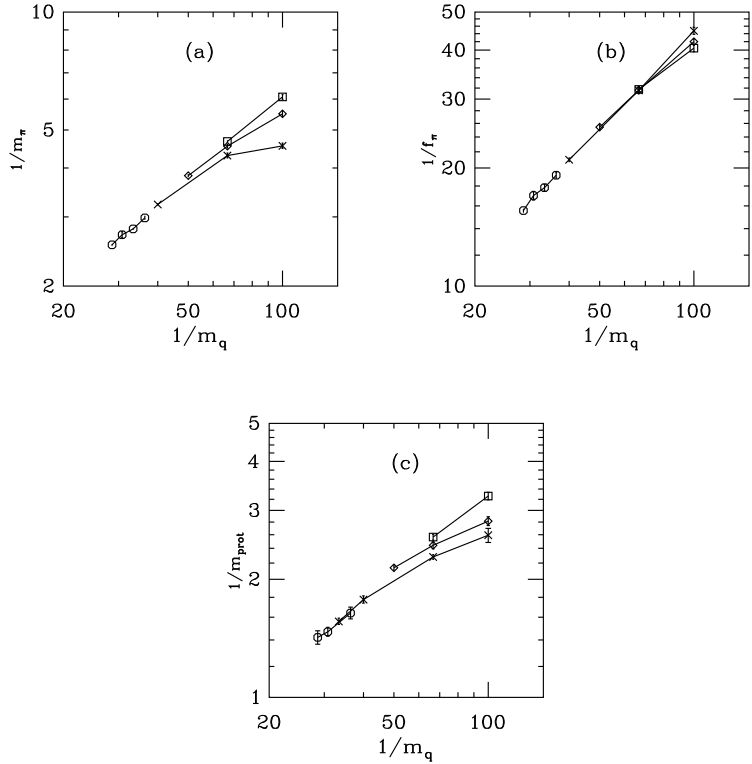}
\end{center}
\caption{Correlation length  versus inverse
quark mass. 
Panels use the inverse pion mass (a), $f_\pi$ (b), and proton mass (c).
Plotting symbols are for different simulation volumes,
squares, $L=48$;
diamonds, $L=40$;
crosses, $L=32$;
octagons, $L=24$.
\label{fig:xivsm}}
\end{figure}

To begin the analysis, we have to see if the data shows the appropriate qualitative behavior:
does $\xi_L$ seem to flatten out at small $m_q$, at an $L$ - dependent value?
Fig.~\ref{fig:xivsm} shows that  (with one exception) 
the trend is as expected, and furthermore larger $\xi_L$
correlates with larger $L$. 
 Data at smaller masses on larger volumes
 would be desirable (was that not always so?) to push to the $m_q$ independent regime,
 but in addition, smaller volume data would do as well.
The size of the finite volume effect is, not surprisingly, different for
different observables.

The exception is $f_\pi$, panel (b), which has a tiny
$L$ dependence with an unexpected
order, $\xi_L$ falls with $L$, and no plateau yet observed, as $f_\pi$ always monotonically
decreases with $m_q$. The axial vector matrix element $\langle 0|A_0|\pi\rangle \sim m_\pi
f_\pi$
does approach a plateau because $m_\pi$ does.

By eye, before the different $L$ data separate, the correlation 
lengths seems to show the power law behavior of  Eq.~\ref{eq:corrlen}. Nevertheless,
 the figure illustrates the danger of a simple fit to a power law: if the system is conformal
in the zero mass limit, the smallest
 fermion mass data is presumably closest to conformality, and yet it is the most contaminated by
 finite volume effects. While the data does not show this, presumably
the largest masses could be
 far enough away from the critical region that they might be outside it, following some
different scaling law.
One is then forced to consider cutting the data from both the high and low mass ends, 
to produce a fit to Eq.~\ref{eq:corrlen},
not a desirable procedure.

\begin{figure}
\begin{center}
\includegraphics[width=0.8\textwidth,clip]{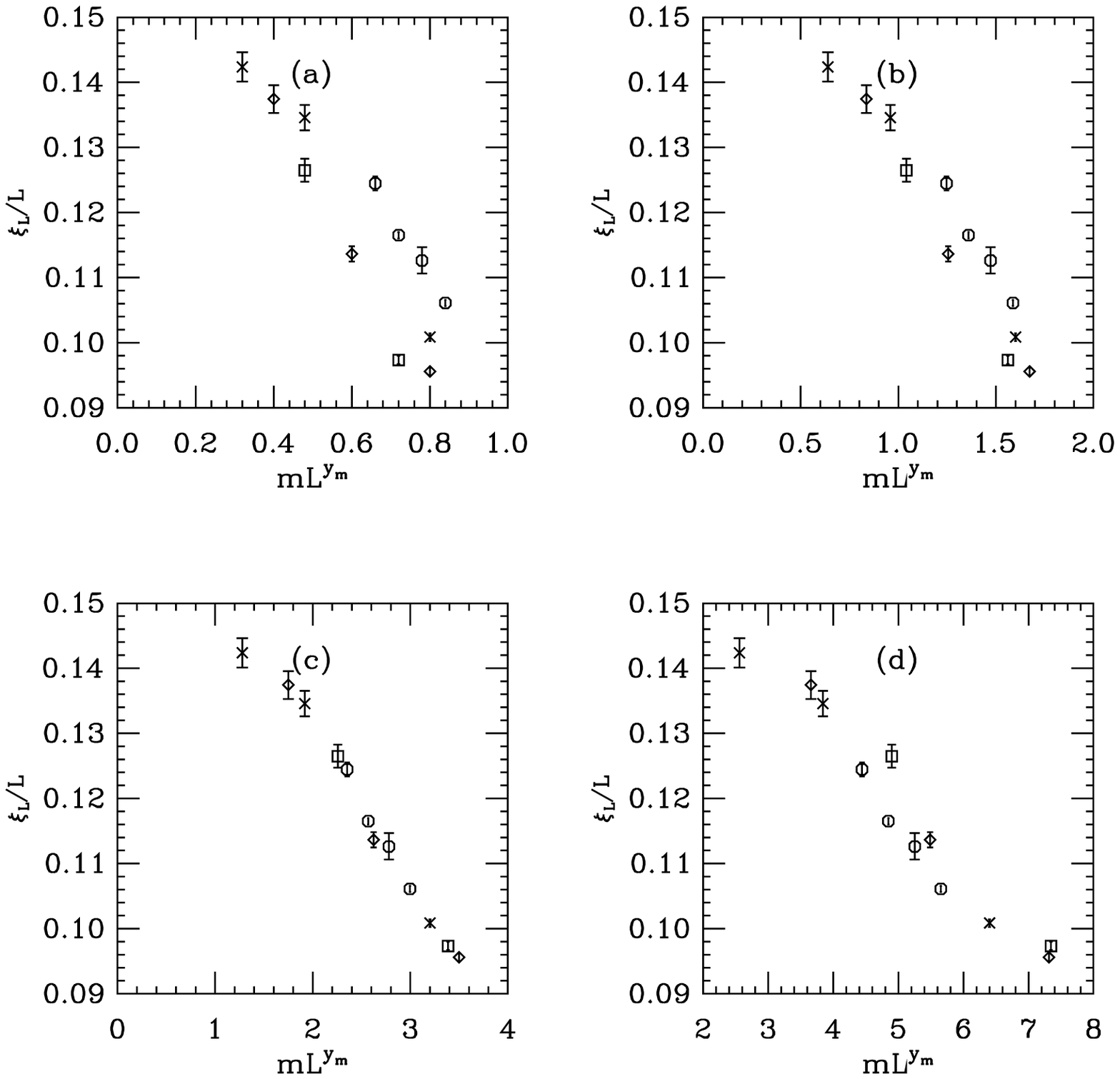}
\end{center}
\caption{
Plots of $\xi_L/L$ vs $m_q L^{y_m}$ with $\xi_L=1/m_\pi$ for four choices of $y_m$:
(a) $y_m=1.0$, (b) $y_m=1.2$, (c) $y_m=1.4$ (d) $y_m=1.6$.
Plotting symbols are for different spatial sizes,
squares, $L=48$;
diamonds, $L=40$;
crosses, $L=32$;
octagons, $L=24$.
\label{fig:allpitest}}
\end{figure}

Next we perform a scan of  $\xi_L/L$ vs $L^{y_m} m_q$, varying $y_m$,
 using the cleanest data set,
 the pseudoscalar. This is shown in Fig.~\ref{fig:allpitest}, for $y_m=1.0$,
1.2, 1.4, 1.6. Collapse to a single scaling curve seems to be occurring.
Notice that $\xi_L/L$ is not zero; the volume is not infinite.

Finally, I attempt to determine a best fit value of $y_m$. The method is that of
Bhattacharjee and Seno \cite{BhS}. The idea is to use each data set
(each different $L$ values, for the same channel)
 to estimate the scaling curve and to find the $y_m$ which pulls the
other $L$ sets onto it. This is done inclusively; all data sets
take a turn at being the fiducial. 
The quantity to minimize is
\bee
P(y_m) = \frac{1}{N_{over}} \sum_p \sum_{j\ne p} \sum_{i,over}
\left( \frac{\xi_L(m_{i,j})}{L_j} - f_p(L_j^{y_m}m_{i,j})\right)^2
\ee
Data set $p$ is used to
estimate the scaling function $f(x)$. This is done by interpolation, either by polynomials or
rational functions, using the recorded values of $\xi_L/L$. The label ``over'' indicates
that the sum only includes data from set $j$ whose $x$ values, $L_j^{y_m}m_{i,j}$,
overlap the range of $x$'s of set $p$.  The overall
factor
of $1/N_{over}$ counts the total number of points and guards against recording a zero value
of $P$ if there are no overlap.  $P$ is minimized by the optimal $y_m$.
This is folded into a jackknife.

\begin{figure}
\begin{center}
\includegraphics[width=0.8\textwidth,clip]{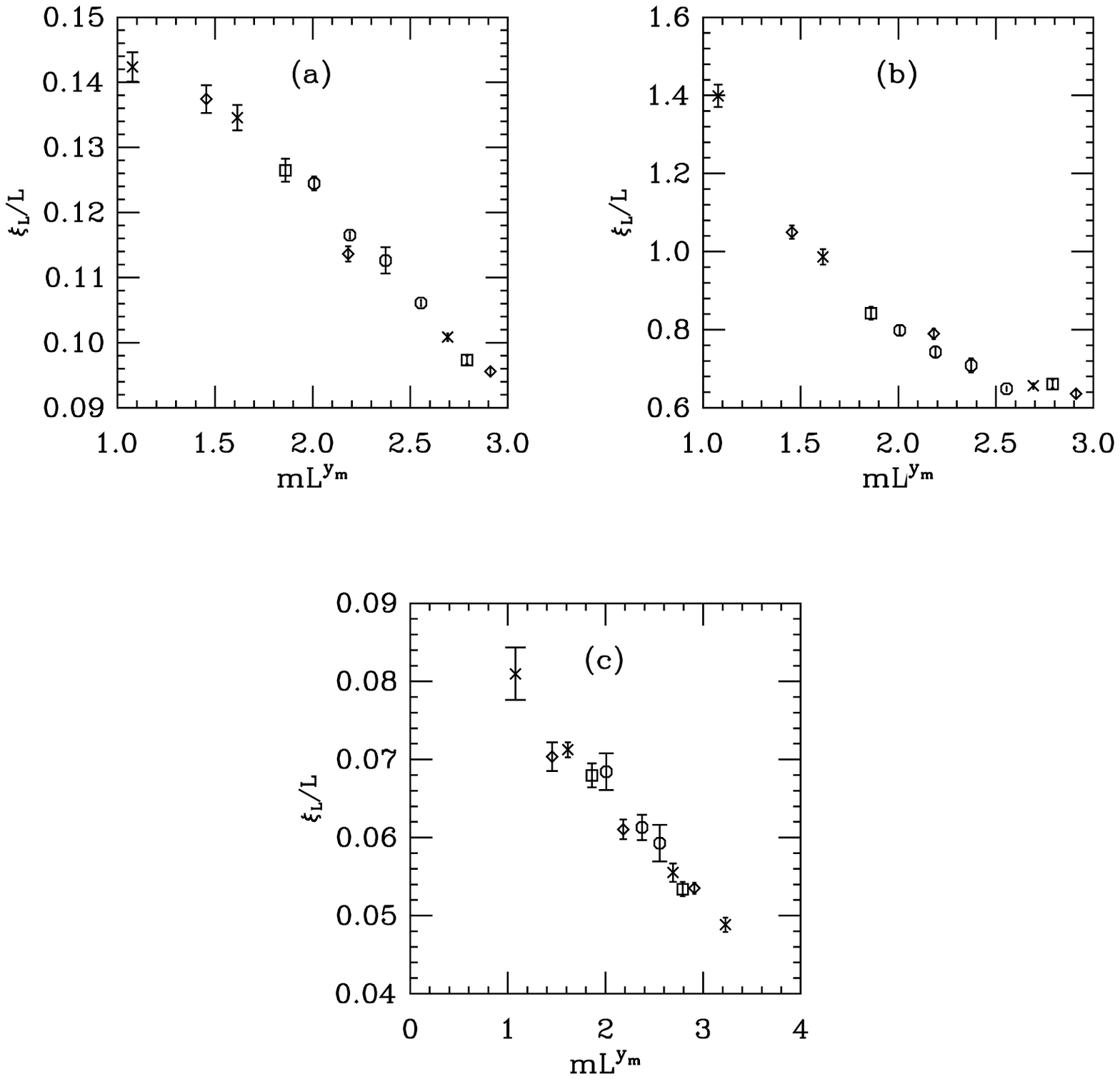}
\end{center}
\caption{Plots of $\xi_L/L$ vs $m_q L^{y_m}$  for  $y_m=1.35$:
(a) pseudoscalar (would-be Goldstone) (b) $f_\pi$, (c) proton.
Plotting symbols are for different spatial sizes,
squares, $L=48$;
diamonds, $L=40$;
crosses, $L=32$;
octagons, $L=24$.
\label{fig:combi}}
\end{figure}

Again, the scaling function is not identical for
different particle correlators, and so I choose not to combine the data further.
All data sets tested so far produce similar values for $y_m$, about 1.35.
Results are given in Table \ref{tab:ym}, with errors from a single-elimination jackknife.
Bhattacharjee and Seno advocate taking an error from an approximation to the second
derivative of P,
\bee
\Delta y_m = \eta y_m\left( 2 \ln \frac{P(y_m(1+\eta))}{P(y_m)}\right)^{-1/2}
\ee
This gives slightly smaller uncertainty estimates, about 0.1 at $\eta=1$.
 The scaling curves for this $y_m$ are
 shown in Fig.~\ref{fig:combi}.

\begin{table}[t]
\caption{ Exponent $y_m$ from various hadronic channels.
Errors are from a single-elimination jackknife. \label{tab:ym}}
\begin{ruledtabular}
\begin{tabular}{cccc}
channel & $y_m$ \\
\hline
pseudoscalar & 1.35(23) \\
nucleon      & 1.43(26) \\
$f_\pi$      &  1.23(31) \\
vector       &  1.33(22) \\
axial vector &  1.32(12) \\
\end{tabular}
\end{ruledtabular}
\end{table}

The authors of Ref.~\cite{Fodor:2011tu} confronted their data with Eq.~\ref{eq:corrlen}
and concluded that the  hypothesis was disfavored. They repeated their analysis
using Eq.~\ref{eq:fss1},  making a specific choice for $F(x)$, and reached a similar conclusion.
The choice of a particular functional form for $F(L/\xi)$ is, I have already remarked, unjustified,
and the authors' analysis may simply show that their functional form for $F(x)$ is not
the one which actually describes the data.

My analysis of the data of Ref.~\cite{Fodor:2011tu} with the assumption that the infinite
volume theory is conformal while the conformality is broken by both the quark mass
 and the finite volume produces a consistent picture, that the leading relevant exponent
at their simulation parameters is $y_m=1.35$ or $\gamma_m=0.35$, with unfortunately
unimpressively large uncertainty. Of course, this is an analysis for which the data set was not designed.
It could be improved by more mass values at all chosen volumes.

 The small $\gamma_m$ measured here
resembles results from other nearly-conformal theories observed to date
\cite{Bursa:2009we,Bursa:2010xn,DeGrand:2010na,DeGrand:2011qd}.

Notice, finally, that
 this is far from being a complete story. For the fit itself, one could be concerned with,
and include, non-scaling contributions. (See  Ref.~\cite{Appelquist:2011dp} which
does this. The authors specified their $F(x)$ rather than letting the data do so.)

More importantly, the $y_m$ which comes out is very likely
not to be an actual scaling exponent.
A few moment's reflection shows why:
 Because the gauge coupling runs so slowly,
simulations done over a small range of volumes
 cannot flow to a fixed point (if it exists) unless they begin very close to it.

 That this is expected,
is easy to see from the one-loop beta function result, where under a scale change $s$
the inverse coupling is shifted by
\bee
\frac{1}{g^2(s)} - \frac{1}{g^2(1)} \sim \frac{b_1}{8\pi^2}\log s + \dots
\ee
In three-flavor $SU(3)$, $b_1=9$ and in 12 flavor $SU(3)$, $b_1=3$, so
 that the equivalent scale changes in the two theories, for an equal change in coupling,
 are $s_{12}=s_3^3$. In ordinary QCD, the coupling runs from weak at a distance
of 0.1 fm to strong at a distance 1.0 fm, or over $s=10$.
 It is hard to get a large aspect ratio $s$ from a set of
numerical simulations at a single set of bare parameters. Therefore,
whether or not a running coupling in one of these many-fermion
theories actually has an IR fixed point, it runs so slowly that
for all practical purposes its running can be neglected. Then the zero mass limit is
effectively conformal. That is all that is needed to 
motivate a scaling analysis as is done here.

 (The lowest order beta function argument is merely suggestive
of problems, but if the theory has an IRFP, than the situation on the weak coupling
side calls for even slower running than the one loop formula.)

Taking this argument further, it says that a measurement of $y_m$ at one set of bare coupling
values does not answer the question of whether the theory actually has an IRFP.
Simulations which have made predictions for an exponent
 map out $g^2$ in some prescription and $\gamma_m(g^2)$, both from simulations at
 many values of the bare parameters. They
then separately determine the 
critical coupling (in some scheme) $g_*^2$ and read off $\gamma_m(g_*^2)$.
(Compare Refs.~\cite{Bursa:2009we,Bursa:2010xn,DeGrand:2010na,DeGrand:2011qd}.)
 Most likely, finite size scaling studies of spectroscopy are just  not the method of choice for
discovering whether the theory has an IRFP, and if it does, accurately determining
either $g_*^2$ or $\gamma_m(g_*^2)$.

\begin{acknowledgments}
%%%%%%%%%%%%%%%%%%%%%%%%%%%%%%%%%%%%%%%%%%%%%%%%%%%%%%%%%%%%%%%%%%%%%%
I thank A.~Hasenfratz, Y.~Shamir, and B.~Svetitsky
 for discussions, C.~Schroeder for correspondence and D.Schaich for a careful reading of the
manuscript. I am grateful to the authors of Ref.~\cite{Fodor:2011tu}
for publishing tables of their data.
This work was supported in part by the US Department of Energy.
%
%%%%%%%%%%%%%%%%%%%%%%%%%%%%%%%%%%%%%%%%%%%%%%%%%%%%%%%%%%%%%%%%%%%%
\end{acknowledgments}
%%%%%%%%%%%%%%%%%%%%%%%%%%%%%%%%%%%%%%%%%%%%%%%%%%%%%%%%%%%%%%%%%%%%%


\begin{thebibliography}{99}
%%%%%%%%%%%%%%%%%%%%%%%%%%%%%%%%%%%%%%%%%%%%%%%%%%%%%%%%%%%%%%%%%%%%%
%\cite{Fodor:2011tu}
\bibitem{Fodor:2011tu}
  Z.~Fodor, K.~Holland, J.~Kuti, D.~Nogradi, C.~Schroeder,
  %``Twelve massless flavors and three colors below the conformal window,''
  Phys.\ Lett.\  {\bf B703}, 348-358 (2011).
    [arXiv:1104.3124 [hep-lat]].

%\cite{Appelquist:2007hu}
\bibitem{Appelquist:2007hu}
  T.~Appelquist, G.~T.~Fleming and E.~T.~Neil,
  %``Lattice study of the conformal window in QCD-like theories,''
  Phys.\ Rev.\ Lett.\  {\bf 100}, 171607 (2008).
  %[Erratum-ibid.\  {\bf 102}, 149902 (2009)]
  %[arXiv:0712.0609 [hep-ph]].
  %%CITATION = PRLTA,100,171607;%%

%\cite{Appelquist:2009ty}
\bibitem{Appelquist:2009ty}
  T.~Appelquist, G.~T.~Fleming and E.~T.~Neil,
  %``Lattice Study of Conformal Behavior in SU(3) Yang-Mills Theories,''
  Phys.\ Rev.\  D {\bf 79}, 076010 (2009),
  %[arXiv:0901.3766 [hep-ph]].
  %%CITATION = PHRVA,D79,076010;%%

%%%%%%%%%%%%%%%%%%%%%%%%%%% More Nf=12 (?) %%%%%%%%%%%%%%%%%%%%%%%



%\cite{Deuzeman:2008sc}
\bibitem{Deuzeman:2008sc}
  A.~Deuzeman, M.~P.~Lombardo, E.~Pallante,
  %``The Physics of eight flavours,''
  Phys.\ Lett.\  {\bf B670}, 41-48 (2008).
  %[arXiv:0804.2905 [hep-lat]].
%FAVORS CHSB, 8 Flavors

%\cite{Fodor:2009wk}
\bibitem{Fodor:2009wk}
  Z.~Fodor, K.~Holland, J.~Kuti, D.~Nogradi and C.~Schroeder,
  %``Nearly conformal gauge theories in finite volume,''
  Phys.\ Lett.\  B {\bf 681}, 353 (2009).
  %[arXiv:0907.4562 [hep-lat]].
  %%CITATION = PHLTA,B681,353;%%
%Nf=4,8,9 CHIRALLY BROKEN


%\cite{Deuzeman:2009mh} %12 is conformal
\bibitem{Deuzeman:2009mh}
  A.~Deuzeman, M.~P.~Lombardo, E.~Pallante,
  %``Evidence for a conformal phase in SU(N) gauge theories,''
  Phys.\ Rev.\  {\bf D82}, 074503 (2010).
  %[arXiv:0904.4662 [hep-ph]].
%FAVORS CONFORMAL


%\cite{Hasenfratz:2009ea}
\bibitem{Hasenfratz:2009ea}
  A.~Hasenfratz,
  %``Investigating the critical properties of beyond-QCD theories using Monte arlo Renormalization Group matching,''
  Phys.\ Rev.\  {\bf D80}, 034505 (2009).
  %[arXiv:0907.0919 [hep-lat]].
% NO Nf=12


%\cite{Jin:2009mc} %12 chirally broken
\bibitem{Jin:2009mc}
  X.~-Y.~Jin, R.~D.~Mawhinney,
  %``Lattice QCD with 8 and 12 degenerate quark flavors,''
  PoS {\bf LAT2009}, 049 (2009).
  %[arXiv:0910.3216 [hep-lat]].
% FAVORS CHSB

%\cite{Hasenfratz:2009kz} % 12 unclear 
\bibitem{Hasenfratz:2009kz}
  A.~Hasenfratz,
  %``Scaling properties of many-fermion systems from MCRG studies,''
  PoS {\bf LAT2009}, 052 (2009)
  [arXiv:0911.0646 [hep-lat]].
  %%CITATION = POSCI,LAT2009,052;%%
% UNKNOWN

%\cite{Hasenfratz:2010fi} %12 conformal
\bibitem{Hasenfratz:2010fi}
  A.~Hasenfratz,
  %``Conformal or Walking? Monte Carlo renormalization group studies of SU(3) gauge models with fundamental fermions,''
  Phys.\ Rev.\  {\bf D82}, 014506 (2010).
  %[arXiv:1004.1004 [hep-lat]].
% FAVORS CONFORMAL

%\cite{Deuzeman:2010fn}
\bibitem{Deuzeman:2010fn}
  A.~Deuzeman, E.~Pallante and M.~P.~Lombardo,
  %``The Bulk transition of many-flavour QCD and the search for a UVFP at strong
  %coupling,''
  PoS {\bf LATTICE2010}, 067 (2010)
  [arXiv:1012.5971 [hep-lat]].
  %%CITATION = POSCI,LATTICE2010,067;%%
%FAVORS CONFORMAL

%\cite{Deuzeman:2010gb} %1/y_m=1.64(5) and 1.52(11)
\bibitem{Deuzeman:2010gb}
  A.~Deuzeman, M.~P.~Lombardo and E.~Pallante,
  %``Chiral symmetry of QCD with twelve light flavors,''
  arXiv:1012.6023 [hep-lat].
  %%CITATION = ARXIV:1012.6023;%%
%FAVORS CHIRALLY RESTORED

%\cite{Hasenfratz:2011xn}
\bibitem{Hasenfratz:2011xn}
  A.~Hasenfratz,
  %``Infrared fixed point of the 12-fermion SU(3) gauge model based on 2-lattice
  %MCRG matching,''
  arXiv:1106.5293 [hep-lat].
  %%CITATION = ARXIV:1106.5293;%%
% FAVORS CONFORMAL
%%%%%%%%%%%%%%%%%%%%%%%%%%%%%%%%%%%%%%%%%%%%%%


%\cite{DeGrand:2009hu}
\bibitem{DeGrand:2009hu}
  T.~DeGrand,
  %``Finite-size scaling tests for SU(3) lattice gauge theory with color sextet fermions,''
  Phys.\ Rev.\  {\bf D80}, 114507 (2009).
  [arXiv:0910.3072 [hep-lat]].

\bibitem{DeGrand:2010na}
  T.~DeGrand, Y.~Shamir and B.~Svetitsky,
  %``Running coupling and mass anomalous dimension of SU(3) gauge theory with
  %two flavors of symmetric-representation fermions,''
  Phys.\ Rev.\  D {\bf 82}, 054503 (2010)
  [arXiv:1006.0707 [hep-lat]].
  %%CITATION = PHRVA,D82,054503;%%


%\cite{BhS}
\bibitem{BhS}
S.~M.~Bhattacharjee and F.~Seno,
%``A measure of data collapse for scaling''
J.\ Phys. \ A: Math. Gen. {\bf 34}, 6375 (2001).

%\cite{Bursa:2009we}
\bibitem{Bursa:2009we}
  F.~Bursa, L.~Del Debbio, L.~Keegan, C.~Pica and T.~Pickup,
  %``Mass anomalous dimension in SU(2) with two adjoint fermions,''
  Phys.\ Rev.\  D {\bf 81}, 014505 (2010)
  [arXiv:0910.4535 [hep-ph]].
  %%CITATION = PHRVA,D81,014505;%%

%\cite{Bursa:2010xn}
\bibitem{Bursa:2010xn}
  F.~Bursa, L.~Del Debbio, L.~Keegan, C.~Pica and T.~Pickup,
  %``Mass anomalous dimension in SU(2) with six fundamental fermions,''
  Phys.\ Lett.\  B {\bf 696}, 374 (2011)
  [arXiv:1007.3067 [hep-ph]].
  %%CITATION = PHLTA,B696,374;%%

%\cite{DeGrand:2011qd}
\bibitem{DeGrand:2011qd}
  T.~DeGrand, Y.~Shamir and B.~Svetitsky,
  %``Infrared fixed point in SU(2) gauge theory with adjoint fermions,''
  arXiv:1102.2843 [hep-lat].
  %%CITATION = ARXIV:1102.2843;%%

%%%%%%%%%%%%%%%%%%%%%%%%%%%%%%%%%%%%% others %%%%%%%%%%%%%%%%%%%%%%%%

%\cite{Appelquist:2011dp}
\bibitem{Appelquist:2011dp}
  T.~Appelquist, G.~T.~Fleming, M.~Lin, E.~T.~Neil and D.~A.~Schaich,
  %``Lattice Simulations and Infrared Conformality,''
  Phys.\ Rev.\  D {\bf 8r41}, 054501 (2011)
  arXiv:1106.2148 [hep-lat].
  %%CITATION = ARXIV:1106.2148;%%


\end{thebibliography}
\end{document}